# First order transition in $Pb_{10-x}Cu_x(PO_4)_6O$ (0.9<x<1.1) containing $Cu_2S$

Shilin Zhu, Wei Wu[#], Zheng Li, Jianlin Luo[$]

*Beijing National Laboratory for Condensed Matter Physics and Institute of Physics,*

*Chinese Academy of Sciences, Beijing 100190, China*

**ABSTRACT**

Lee et al. reported that the compound LK-99, with a chemical formula of $Pb_{10-x}Cu_x(PO_4)_6O$ (0.9<x<1.1), exhibits room-temperature superconductivity under ambient pressure. In this study, we investigated the transport and magnetic properties of pure $Cu_2S$ and LK-99 containing $Cu_2S$. We observed a sharp superconducting-like transition and a thermal hysteresis behavior in the resistivity and magnetic susceptibility. However, we did not observe zero-resistivity below the transition temperature. We argue that the so-called superconducting behavior in LK-99 is most likely due to a reduction in resistivity caused by the first order structural phase transition of $Cu_2S$ at around 385 K, from the β phase at high temperature to the γ phase at low temperature.

**INTRODUCTION**

Superconductors exhibit zero-resistivity and perfect diamagnetism, making them important in applications such as nuclear magnetic resonance, maglev trains, superconducting quantum computing, and lossless power transmission, etc. However, the superconducting transition temperatures of known superconductors at ambient pressure are relatively low. For example, the highest superconducting transition temperature $T_c$ observed in high-temperature cuprates is only around 135 K [1, 2]. The hydrogen-rich materials have discovered under high pressure with $T_c$ approaching room-temperature [3, 4], but the extreme high pressure required makes these materials impractical for applications. Therefore, the exploration of room-temperature superconductors at ambient pressure is the central focus of superconducting material research and the dream of researchers in the field.

Recently, Lee et al. reported a material called LK-99, a modified-lead apatite crystal structure with the composition $Pb_{10-x}Cu_x(PO_4)_6O$ (0.9<x<1.1), to be the first room-temperature superconductor at ambient pressure [5, 6]. They found that LK-99 exhibited a sharp drop in resistance to zero and diamagnetism around 378 K. This report immediately attracted widespread attention, and a number of research groups attempted to verify the superconductivity of LK-99 [7-10]. Several groups came up with different conclusions and none confirmed room-temperature superconductivity.

We notice that the reported LK-99 sample contained a certain amount of $Cu_2S$ impurity [5, 8, 9, 10], which undergoes a structural phase transition from the β phase at high temperature to the γ phase at low temperature near 400 K [11, 12]. To investigate whether the superconducting-like transition observed is intrinsic for LK-99 or caused by the $Cu_2S$ impurity, we studied the transport and magnetic properties of $Cu_2S$ and the mixture of LK-99 and $Cu_2S$. We found that the resistivity of $Cu_2S$ decreased by 3 - 4 orders of magnitude around 385 K, close to the reported transition temperature and resistance behavior in references [5, 6]. Additionally, we measured the resistivity of the mixture of LK-99 and $Cu_2S$, which show a sharp resistivity transition at the temperature consistent with the reported findings, but without zero resistance. Based on our measurements of resistivity and magnetization, we argue that the superconducting-like behavior in LK-99 most likely originates from the first-order structural phase transition of $Cu_2S$.

**RESULTS AND DISCUSSION**

$Cu_2S$ powder was commercially available (Alfa Aesar) (99.5% −325 mesh). The preparation procedure for LK-99 follows the methodology reported in reference [5, 6]. The precursor $Cu_3P$ and Lanarkite of LK-99 were prepared separately by solid-state reactions. To synthesize $Cu_3P$, Cu powder (Alfa Aesar 99.99%) was mixed with P (Aladdin 99.999%) in a molar ratio of 3:1 inside a glovebox. The mixture was then pressed into a pellet and sealed in high-vacuum quartz tubes. The tubes were heated to 300 °C at a rate of 4 °C/min, and kept at that temperature for 20 hours. Subsequently, the temperature is raised to 900 °C for 10 hours. Finally, the sample was kept at 900 °C for 10 hours. Lanarkite was obtained by annealing a stoichiometric mixture of commercially available $Pb(SO)_4$ powder (Aladdin 99.99%) and PbO powder (Aladdin 99.999%) to 725 °C at a rate of 4 °C/min for 24 hours. (Annealed in the air and in the vacuum will affect $Cu_2S$ content in the final products.) Polycrystalline samples of LK-99 were prepared by solid state reactions of $Cu_3P$ powder and Lanarkite $Pb_2(SO_4)O$ powder in a 1:1 stoichiometric ratio. The resulting powders were pressed into pellets and sealed into evacuated fused silica ampoules. The ampoules were annealed at 925 °C (heating rate: 5 °C/min) for 24 hours in a box furnace. After cooling to room temperature, the samples were ground in argon atmosphere, pressed into pellets, and then annealed again at 925 °C for 24 hours.

All products were structurally characterized using x-ray powder diffraction (Rigaku) in the Bragg−Brentano geometry, equipped with a Cu Kα x-ray source (λ = 1.5406 Å). Rietveld refinements were performed using TOPAS software. The resistivity was measured using a standard 4-probe method between 2 K and 400 K in PPMS system of Quantum Design Company. To measure transport properties, $Cu_2S$ powder was loaded into a metallic die and pressed into a dense pellet. The magnetic susceptibility was measured in temperature range of 2 K - 400 K using a SQUID VSM Magnetometer of Quantum Design Company.

We synthesized two kinds of LK-99 with different $Cu_2S$ content, using different precursor of Lanarkite annealed in the air and in the vacuum, respectively. Both $Cu_2S$ and LK-99 phases are observed from the x-ray diffraction (XRD) pattern of the mixtures. Figure 1(a) and (b) shows the XRD patterns of pure $Cu_2S$ and two mixtures of LK-99 and $Cu_2S$. The XRD pattern of $Cu_2S$ phases are well indexed on the basis of monoclinic-type structure with the space group $P2_1/c$. The LK-99 match the indexed of database which contain impurity phase of $Cu_2S$. The intensity ratio reflect the $Cu_2S$ content is approximate 5% in sample 1 (S1) and 70% in sample 2 (S2). Figure 1(c) shows the main peaks of $Cu_2S$ in pure $Cu_2S$ and S1 and S2 between 45° and 51°. We notice a slight displacement of the peak position for $Cu_2S$ in the mixture compared with pure $Cu_2S$, which may be attributed to the difference in the S content of $Cu_2S$.

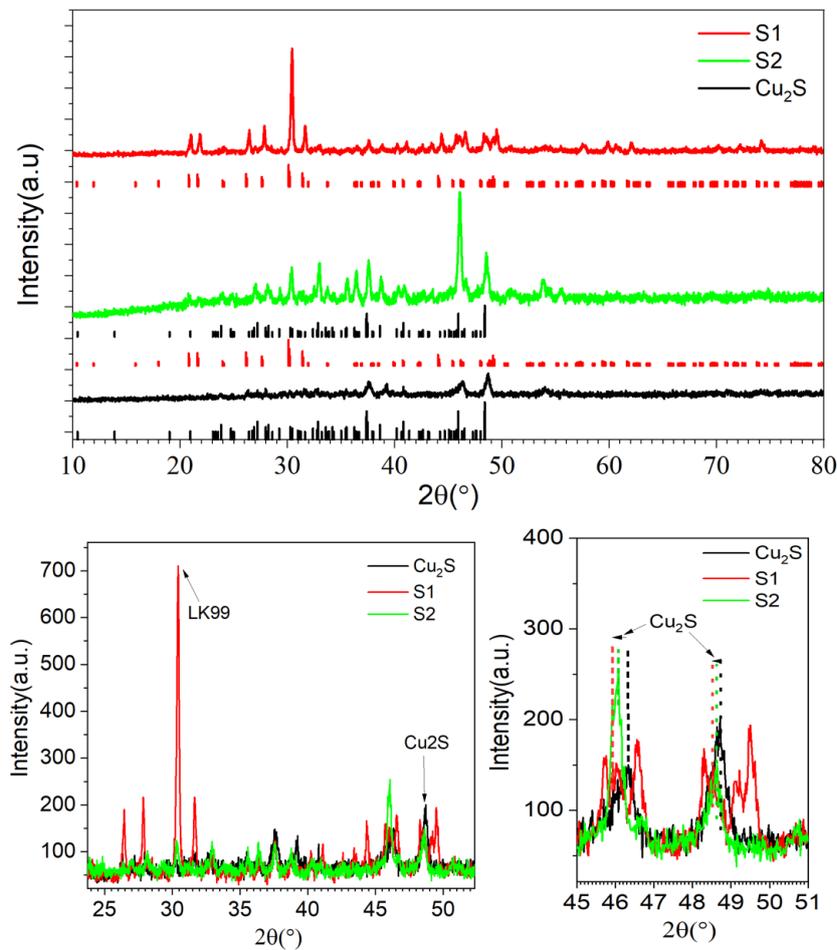

Fig. 1 (a) XRD patterns of pure $Cu_2S$ and S1 and S2. (b) The main XRD peaks of $Cu_2S$ in pure $Cu_2S$ and S1 and S2 (c) The enlarge view of (b) in the range between 45° and 51°.

Figure 2 shows the resistivity of $Cu_2S$ in a temperature range of 2 K - 400 K,

plotted on both linear (Fig. 2(a)) and logarithmic scales (Fig. 2(b)), respectively. At around 385 K, a significant decrease in the resistivity of Cu$_2$S is observed, with a reduction from 13.7 Ohm cm at 385.6 K to 0.006 Ohm cm at 381.5 K, representing a drop of 3 - 4 orders of magnitude. This drop in resistivity is similar to the resistance drop observed in LK-99, as reported by Lee et al., where the drop temperature is around 378 K [5, 6].

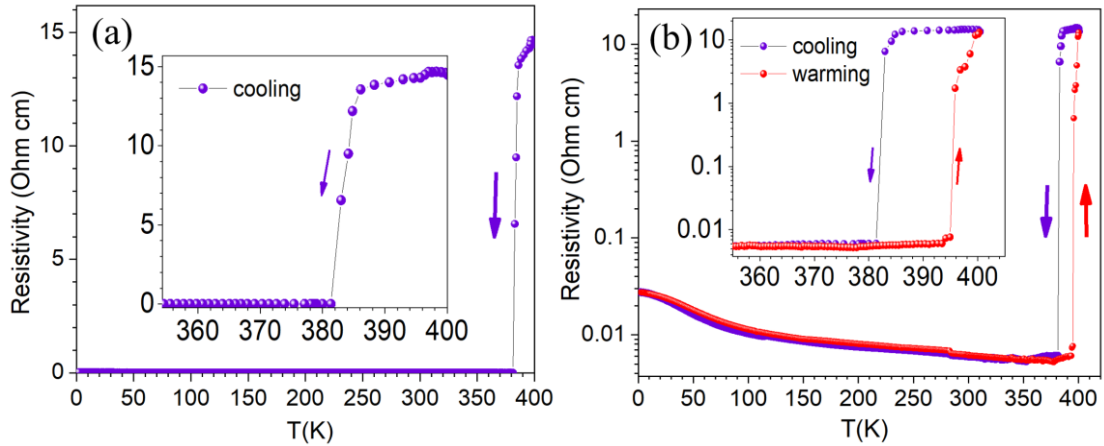

Fig. 2 (a) Temperature dependence of resistivity of Cu$_2$S pallet with a very sharp transition. (b) Temperature dependence of resistivity in cooling and warming course in logarithmic coordinates of Cu$_2$S with a wide hysteresis behavior.

Cu$_2$S undergoes a structural phase transition from a hexagonal structure at high temperatures to a monoclinic structure at low temperatures near 400 K. This transition has been observed in conductivity and coefficient of thermal expansion measurements [11, 12]. Figure 2(b) shows the resistivity of Cu$_2$S maintains a finite non-zero value below the transition temperature, and gradually increases as the temperature decreases. Additionally, a wide thermal hysteresis of 13.1 K is found in the resistivity of Cu$_2$S when comparing the cooling curve and warming curve, indicating a first-order phase transition.

Given the sudden drop in resistivity by 3-4 orders of magnitude and the similarities in the near-zero resistance and phase transition temperature with LK-99 [5, 6], it strongly suggests that the superconductivity-like behavior in LK-99 reported by Lee et al. is caused by the structural phase transition of the impurity Cu$_2$S. In order to verify our hypothesis, we conducted resistivity measurements on the mixture of LK-99 with different content of Cu$_2$S, as shown in Fig. 3.

Figure 3(a) and (b) shows a sharp drop in resistivity around 370 K of S2, accompanied by a pronounced thermal hysteresis. As the temperature decreases, the resistivity displays metallic behavior (d$\rho$/dT > 0) over a wide high temperature range.

Below 100 K, the resistivity increases with decreasing temperature, showing a semiconducting-like behavior. The sharp drop in resistivity and the transition temperature are similar to that observed by Lee et al. [5, 6]. Figure 3(c) and (d) shows temperature dependence of resistivity of S1. We observed a jump in resistivity around 370 K, thermal hysteresis behavior also existed in this sample.

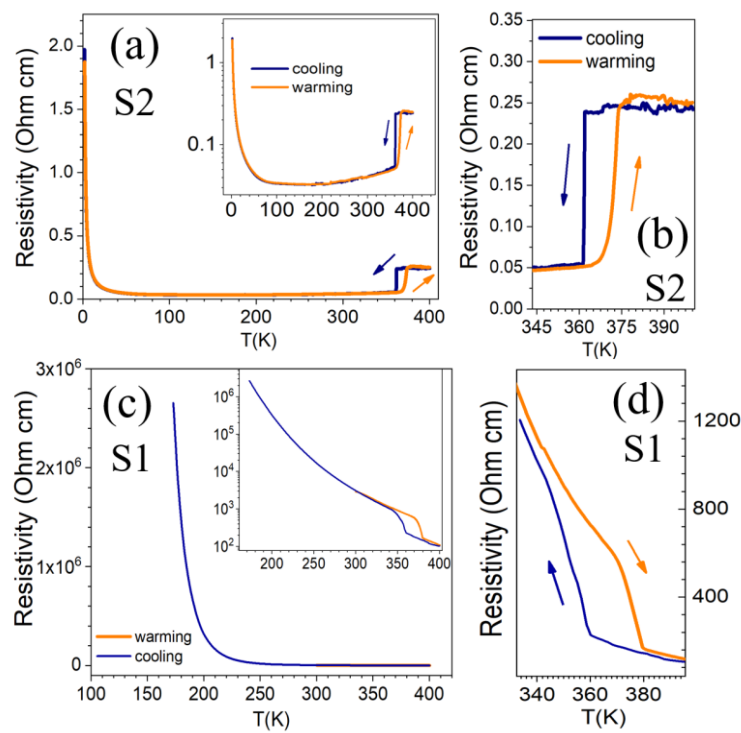

Fig. 3 (a)Temperature dependence of resistivity of S2. The inset is shown in logarithmic coordinates (b) The enlarge view of the transition loop of S2. (c) Temperature dependence of resistivity of S1. The inset is shown in logarithmic coordinates. (d) The enlarge view of the transition loop of S1.

Figure 4(a) shows the magnetic susceptibility of S2 in a temperature range of 2 K - 400 K, in a field of 1 T. The magnetic susceptibility exhibits diamagnetism with a phase transition occurring at 375 K. Once again, a pounced thermal hysteresis of more than 10 K is observed in the magnetic susceptibility data, suggesting that there is a first-order phase transition. Additionally, the transition temperature range closely aligned with the structural phase transition temperature of $Cu_2S$. Figure 4(b) shows the magnetization as a function of magnetic field, ranging from -7 T to 7 T at various temperatures from 2 K to 400 K. It exhibits typical diamagnetic behavior, with no resemblance to the behavior of a superconductor.

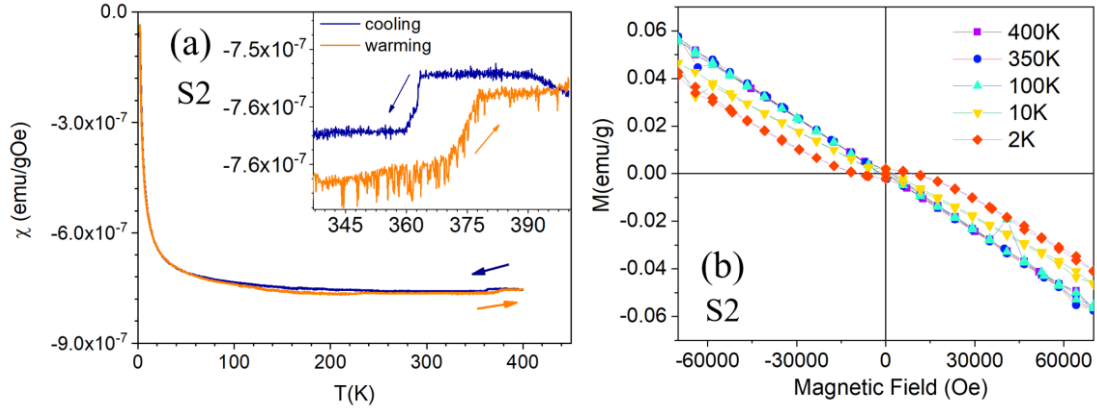

Fig. 4 (a) Temperature dependence of diamagnetic susceptibilities measured in S2 at 1 T in cooling and warming course. (b) The magnetic field dependence of magnetic moment of S2

The findings above strongly suggest the superconducting-like transition in LK-99 as reported by Lee et al. [5, 6] originates first order structural transition of the impurity phase of $Cu_2S$ from a hexagonal structure in β phase at high temperature to a monoclinic structure in γ phase at low temperature around 385 K. It is important to note that this first-order structural transition differs significantly from the second-order superconducting transition. We recommend that Lee et al. perform resistivity measurements during cooling and warming processes on their superconducting-like materials to determine if there is thermal hysteresis.

**CONCLUSION**

In conclusion, we measured transport and magnetic properties of pure $Cu_2S$ as well as the mixture LK-99/$Cu_2S$, and reproduce the experimental results of resistivity. We found a sharp drop in resistivity, however, none of them show zero-resistivity. The superconducting-like behavior in LK-99 most likely originates from a magnitude reduction in resistivity caused by the first-order structural phase transition of $Cu_2S$.
**Email address:**  [#] welyman@iphy.ac.cn
　　　　　　　　　　[$] jlluo@iphy.ac.cn

**REFRENCES**
[1] Bednorz , J. G. and Müller, K. A. Z. Phys. B 64, 189 (1986).
[2] Gao, L. et al, Superconductivity up to 164 K in $HgBa_2Ca_{m-1}Cu_mO_{2m+2+\delta}$ (m = 1, 2, and 3) under quasihydrostatic pressures. Phys. Rev. B 50, 4260-4263 (1994).
[3] Drozdov, A. P. Eremets, M. I. Troyan, I. A. Ksenofontov, V. & Shylin, S. I. Conventional superconductivity at 203 kelvin at high pressures in the sulfur hydride system, Nature 525, 73 (2015).

[4] Kong, P. et al. Superconductivity up to 243 K in the yttrium-hydrogen system under high pressure. Nat. Commun. 12, 5075 (2021).

[5] Lee, S. Kim, J. H. & Kwon, Y. W. The First Room-Temperature Ambient-Pressure Superconductor. arXiv:2307.12008 (2023).

[6] Lee, S. et al. Superconductor $Pb_{10-x}Cu_x(PO_4)_6O$ showing levitation at room temperature and atmospheric pressure and mechanism. arXiv:2307.12037 (2023).

[7] Wu, H. et al. Successful growth and room temperature ambient-pressure magnetic levitation of LK-99, arXiv:2308.01516 (2023).

[8] Li, L. et al. Semiconducting transport in $Pb_{10-x}Cu_x(PO_4)_6O$ sintered from $Pb_2SO_5$ and $Cu_3P$, arXiv:2307.16802 (2023).

[9] Kumar, K. Karn, N.K. and Awana, V.P.S. Synthesis of possible room temperature superconductor LK-99: $Pb_9Cu(PO_4)_6O$, arXiv:2307.16402 (2023)

[10] Hou, Q. et al. Observation of zero resistance above 100 K in $Pb_{10-x}Cu_x(PO_4)_6O$, arXiv:2308.01192 (2023)

[11] Hirahara E. The Physical Properties of Cuprous Sulfides-Semiconductors; J. Phys. Soc. Jpn. 6, 422-427 (1951).

[12] Nierodaa, P. et al. Thermoelectric properties of $Cu_2S$ obtained by high temperature synthesis and sintered by IHP method, Ceramics International, 46, Part A, 25460 (2020).